\documentclass[aps,epsfig]{revtex4}
\usepackage{graphicx}% Include figure files
\begin{document}
\def\be{\begin{equation}}
\def\ee{\end{equation}}
\def\bfi{\begin{figure}}
\def\efi{\end{figure}}
\def\bea{\begin{eqnarray}}
\def\eea{\end{eqnarray}}

\title{Fluctuation-Dissipation relations far from Equilibrium}

\author{Federico Corberi$^1$, Eugenio Lippiello$^2$ and
Marco Zannetti$^1$}
\affiliation {$^1$Dipartimento di Matematica ed Informatica
via Ponte don Melillo, Universit\`a di Salerno, 84084 Fisciano (SA), Italy
\newline $^2$Dipartimento di Scienze Fisiche, Universit\'a di Napoli
``Federico II'', 80125 Napoli, Italy}

\begin{abstract}

In this Article we review some recent progresses in the field of
non-equilibrium linear response theory. We show how a generalization
of the fluctuation-dissipation theorem can be derived for 
Markov processes, and discuss the Cugliandolo-Kurchan~\cite{Cugliandolo93}
fluctuation dissipation relation for aging systems 
and the theorem by Franz {\it et. al.}~\cite{Franz98}
relating static and dynamic properties. 
We than specialize the subject to phase-ordering systems 
examining the scaling properties of the linear response function
and how these are determined by the behavior of topological defects.
We discuss how the connection between statics and dynamics can be
violated in these systems at the lower critical dimension or 
as due to stochastic instability.

\end{abstract}

\maketitle


\section{Introduction} \label{intro}

The fluctuation-dissipation theorem~\cite{Kubo} (FDT) is one of the fundamental
accomplishments of linear response theory applied to equilibrium systems.
According to the FDT a response function $\chi$, describing 
the effects of a small perturbation exerted on a system,
is linearly related, via the equilibrium temperature $T$, 
to a correlation function $C$ of the
the system in the absence of the perturbation.
In the language of magnetic systems, which we shall adopt in the following, 
one usually considers the application of
an external magnetic field $h$, and $\chi $ is the magnetic susceptibility.

In recent years there has been a considerable interest,
arisen in different fields  
such as turbulent fluids~\cite{Hohenberg89}, disordered, 
glassy~\cite{Cugliandolo93,Cugliandolo97,Berthier00} and aging systems~\cite{Fielding02},
in the generalization
of the results of linear response theory to out of equilibrium systems.
Differently from equilibrium statistical mechanics, where a well
funded and controlled theory is available, there is not nowadays a theorem
of a generality comparable to the FDT for non-equilibrium states.
Nevertheless, some interesting progresses have been done in understanding 
some particular aspects of non-equilibrium linear response theory, 
some of which will be discussed in this paper. 

A first basic question regards the possibility of generalizing the FDT.
Namely, the question is whether also away from equilibrium it is still possible 
to relate the response
function to properties of the unperturbed dynamics, possibly in the form of correlation functions.
A positive answer to this question exists when the time evolution is Markovian
and described by a differential equation of the Langevin type~\cite{CKP}
or for systems described by a master equation~\cite{noialg}. In this case,
the response function is related to correlation functions of the unperturbed
systems, which, however, are not only the correlation $C$ involved in equilibrium.
These results are particularly important since they allow the study of the
response function without considering the perturbed system, which is generally
more complicated.

Once a relation between response and correlations is established, at least in the
restricted framework of Markov processes, the natural question is  
which piece of information, if any, can be learned from it about the non-equilibrium state.
In equilibrium the linear relation between $\chi $ and $C$ is universal, and the coefficient
$d\chi /dC $ entering this relation is $T$. 
In the restricted area of aging systems it has been shown that the $\chi (C)$ relation
still bears an universal character, although weaker than in equilibrium.
This is because the theorem by Franz, Mezard, Parisi and Peliti~\cite{Franz98} connects
$\chi (C)$ to the equilibrium probability distribution of the overlaps
$P_{eq}(q)$ and different statistical mechanical systems can be classified
into few universality classes on the basis of their 
$P_{eq}(q)$ according to the replica symmetry breaking character 
of the ground state~\cite{Ricci99}.   
Moreover, $d\chi /dC$ can be interpreted as an {\it effective } temperature~\cite{Cugliandolo97}.

These results promoted linear response theory as an important tool to investigate
the non-equilibrium behavior or even the structure of equilibrium states of complex
systems, such as spin glasses, which are hard to equilibrate, where $P_{eq}(q)$
can be better inferred from a non-equilibrium measurement of $\chi (C)$.
However, in order for these studies to be sound, a basic understanding of the out of 
equilibrium behavior of the response function is required.
Instead, already at the level of coarsening systems, which can be considered
as the simplest paradigm of aging phenomenon, where 
a satisfactory general analytic description can nowadays
be given by means of exactly solvable models or approximate theories, 
the scaling properties of the response function are non trivial and still far from being
understood. Notably, the statics-dynamics connection
stated by the theorem~\cite{Franz98} is not always fulfilled in coarsening systems.  

In this paper we review some recent progresses in the field of non-equilibrium
linear response theory. The focus is mainly on aging systems and, in particular,
on phase-ordering, which, because of its relative simplicity, is
better suited for a thorough analysis.
The article is organized as follows: Secs.~\ref{basic},~\ref{offgen},
~\ref{flucdisrel} and~\ref{statdingen} are of a general character;
here we fix up the basic definitions,
discuss the generalization of the FDT for Markov processes, introduce
the fluctuation-dissipation relation and review the theorem~\cite{Franz98}
which links statics to dynamics. In Sec.~\ref{phord} some of the
concepts introduced insofar are specialized to the case of phase-ordering kinetics.
After a general description of the dynamics in Sec.~\ref{falling},
the behavior of the response function is reviewed in Sec.~\ref{zerofc}.
In particular,in Sec.~\ref{scalh} the scaling properties of $\chi $ are discussed
and in Sec.~\ref{roughening} it is shown how, in the case of a scalar order
parameter, the exponents can be related to the roughening properties of the interfaces. 
Sec.~\ref{statdinferro}
contains a discussion of how the connection between statics and dynamics
is realized or violated in coarsening systems. Some open problems are
enumerated in Sec.~\ref{concl} and the conclusions are drawn. 

\section{Basic Definitions} \label{basic}

Let us consider a system described by the Hamiltonian $H_0$.
The autocorrelation function of a generic observable ${\cal O}$ between
the two times $s$ and $t\geq s$ is 
\be
C(t,s)=\langle {\cal O}(t){\cal O}(s)\rangle,
\ee
where $\langle \dots \rangle$ is an ensemble average.
Switching on an impulsive perturbation $h(s)$ at time $s$ which changes the 
Hamiltonian ${\cal H}\to {\cal H}+\Delta {\cal H}={\cal H}-h{\cal O}$, 
the linear (impulsive) response function
is given by
\be
R(t,s) =\left .\frac {\partial \langle {\cal O}(t)\rangle}
{\partial h(s)}\right \vert _{h=0}
\ee
The integrated response function, or dynamic susceptibility, is
\be
\chi (t,s)=\int _s ^t R(t,t')dt'
\label{integ}
\ee
and corresponds to the response to a perturbation switched on from $s$ onwards.

In equilibrium, time translation invariance (TTI) holds, so that all
the two time quantities introduced above depend only on the time
difference $\tau =t-s$. The FDT reads
\be
TR(\tau )=-\frac{dC(\tau)}{d\tau},
\label{fdtr}
\ee
where $T$ is the temperature,
or, equivalently, for the integrated response
\be
T\chi (\tau )=C (\tau=0)-C (\tau ).
\label{fdtchi}
\ee

%In the following we will deal with spin models of the Ising type, with
%\be 
%H_0[\sigma ]=-J \sum _{\langle i,j\rangle } \sigma _i\sigma _j 
%\ee
%where the perturbation is usually assumed in the form of an external magnetic field
%\be
%H_p=-\sum _i \sigma _i h_i
%\ee
%Choosing ${\cal O}\equiv \sigma _i$ one has
%\be
%C(t,s)=\langle \sigma _i(t)\sigma _i(s)\rangle
%\ee
%and the perturbation is the applied magnetic field $h_i$.
%Assuming $h_i$ white noise with $\overline {h_i}=0$
%and $\overline h_i h_j\overline =\delta _{ij}h^2$, the integrated
%response function, or zero field cooled
%magnetization (ZFC) is given by 
%\be
%\chi (t,s)=\frac{1}{h^2} 
%\overline {\langle \sigma _i(t)\rangle h_i(s)}
%\ee

\section{Off-equilibrium generalization of the fluctuation dissipation theorem
for Markov processes} \label{offgen}

Consider
a system with an order parameter field $\phi(\vec{x})$ evolving with the Langevin equation of motion
\be
{\partial \phi(\vec{x},t) \over \partial t} = B\left [\phi(\vec{x},t) \right ] + \eta (\vec{x},t)
\label{I1}
\ee
where $B\left [\phi(\vec{x},t)\right ]$ is the deterministic force and $\eta (\vec{x},t)$ is
a white, zero-mean Gaussian noise. In this framework a generalization of the FDT was derived
in~\cite{CKP}. Let us recall the basic elements, referring to~\cite{CKP} for further
details.
From Eq.~(\ref{I1}), the linear response function is simply computed as the 
correlation function of the order parameter with the noise
\be
2T R(t,s) = \langle \phi(\vec{x},t) \eta (\vec{x},s)\rangle
\label{I2}
\ee
where $T$ is the temperature of the thermal bath and $t \geq s$ by causality. 
It is straightforward to
recast the above relation in the form
\be
TR(t,s)=\frac{1}{2}\frac{\partial C(t,s)}{\partial s}
-\frac{1}{2}\frac{\partial C(t,s)}{\partial t} -A(t,s)
\label{primi}
\ee
where 
\be
C(t,s)=\langle\phi(\vec x,t)\phi(\vec x,s)\rangle-
\langle\phi(\vec x,t)\rangle\langle\phi(\vec
x,s)\rangle
\label{prii}
\ee
and
\be
A(t,s)\equiv \frac{1}{2}\left \{
\langle \phi(\vec{x},t) B\left [\phi(\vec{x},s)\right ] \rangle- 
\langle B\left [\phi(\vec{x},t)\right ]\phi(\vec{x},s)\rangle
\right \}
\label{ax}
\ee
is the so called asymmetry.  
Eq.~(\ref{primi}), or~(\ref{I2}),  qualifies as an extension of the FDT out of equilibrium, 
since in the right hand side
there appear unperturbed correlation functions. When time translation and 
time inversion invariance hold, so that $A(t,s)=0$ and $\partial C(t,s)/\partial t=
-\partial C(t,s)/\partial s$, it reduces to
the equilibrium FDT~(\ref{fdtr}). Let us mention that this equation holds~\cite{noialg} 
in the same form both for conserved order parameter (COP) and 
non conserved order parameter (NCOP) dynamics~\cite{Bray94}.

The next interesting question is whether one can do the same also in the case of discrete 
spin variables, where the kinetics is described by a master-equation, there is no 
stochastic differential equation and, therefore, Eq.~(\ref{I2}) is not available.
A first approach to this problem was undertaken in Refs.~\cite{chat,ricci,diez,Crisanti2002}
where a relation between the response function and particular correlators
was obtained. As we shall discuss briefly below, however, their
results cannot be qualified as generalizations of the fluctuation-dissipation
theorem. Instead, in what follows we scketch how (details can be found in~\cite{noialg}), 
an off- equilibrium generalization of
the FDT, which takes {\it exactly the same form} as Eqs.~(\ref{primi},\ref{ax}) 
and which holds, as
in the Langevin case, for NCOP (spin flip) and COP
(spin exchange) dynamics can be derived also in this case. 

Let us consider a system of Ising spins $\sigma_i=\pm 1$ executing a Markovian stochastic process. 
The generalization to $q$-states spins, as in the Potts or Clock model, is straightforward.
The problem is to compute the linear 
response $R(t,s)$ on the spin at the site $i$ and at the time $t$, due to 
an impulse of external field at an earlier time $s$ and at the same site $i$. Let  
\be
h_j(t)=h \delta_{i,j}
\theta (t-s)\theta (s+\Delta t -t)
\label{pert}
\ee
be the magnetic field on the $i$-th site acting during 
the time interval $[s,s+\Delta t]$, where $\theta $ is the Heavyside step function. 
The response function then is given by \cite{chat,Crisanti2002}
\be
R(t,s)=\lim_{\Delta t \to 0} \frac{1}{\Delta t}
\left . \frac{\partial \langle \sigma_i(t) \rangle}{\partial h_j(s)} \right
\vert _{h=0}
\label{4}
\ee
where 
\be
\left .\frac{\partial \langle \sigma_i(t) \rangle}{\partial h_j(s)} \right 
\vert _{h=0}=\sum _{[\sigma],[\sigma'],[\sigma'']}
\sigma_i p([\sigma],t\vert [\sigma'],s+\Delta t) 
\left .\frac{\partial p^h([\sigma'],s+\Delta t \vert [\sigma''],s)}
{\partial h_j} \right \vert _{h=0} p([\sigma''],s) 
\label{5}
\ee
and $[\sigma]$ are spin configurations.

Let us concentrate on  the factor containing the conditional probability in the presence 
of the external field $p^h([\sigma'],s+\Delta t \vert [\sigma''],s)$.
In general, the conditional probability for $\Delta t$ sufficiently small is given by
\be
    p([\sigma],t+\Delta t \vert [\sigma'],t)= \delta_{[\sigma],[\sigma']}+
         w([\sigma']\to [\sigma]) \Delta t + {\cal O}(\Delta t^2),
    \label{pippo}
\ee
where we have used the boundary condition $p([\sigma],t\vert [\sigma'],t)= 
\delta_{[\sigma],[\sigma']}$. 
Furthermore, the transition rates must verify detailed balance
\be
w([\sigma] \to [\sigma'])\exp(-{\cal H}[\sigma]/T)=w([\sigma'] \to [\sigma])
\exp(-{\cal H}[\sigma']/T),
\label{detbal}
\ee
where ${\cal H}[\sigma]$ is the Hamiltonian of the system. 

Introducing the perturbing field as an extra term 
$\Delta {\cal H}[\sigma]=-\sigma_j h_j$ 
in the Hamiltonian, to linear order in $h$ the most general form 
of the perturbed transition rates $w^h([\sigma] \to [\sigma'])$ 
compatible with the detailed balance condition is 
\be
w^h([\sigma] \to
[\sigma'])=w^0([\sigma] \to [\sigma'])
\left  \{1-\frac{1}{2 T}h_{j}(\sigma_{j}-\sigma'_j) + M([\sigma],[\sigma'])
\right  \}, 
\label{transh}
\ee
where $M([\sigma],[\sigma'])$ is an arbitrary 
function of order $h/T$ symmetric with respect to the exchange 
$[\sigma] \leftrightarrow [\sigma']$,  
 and $w^0([\sigma] \to [\sigma'])$ are unspecified unperturbed transition rates, which satisfy
detailed balance. 
In the following, for simplicity, we shall take $M([\sigma],[\sigma'])=0$.
Implication of this choice, which corresponds to a specification of
the perturbed transition rates, are discussed in~\cite{noialg}.

Inserting Eqs.~(\ref{pippo}),~(\ref{transh}) 
in Eq.~(\ref{5}), and
using the time translational invariance of the conditional probability 
$p([\sigma],t\vert[\sigma'],s+\Delta t)=p([\sigma],t-\Delta t\vert[\sigma'],s)$,
after some manipulations the response function~(\ref{4}) can
be written as 
\be
T R(t,s)=\frac{1}{2}\frac {\partial C(t,s)}{\partial s}-
\frac{1}{2}E(t,s)
\label{nuova}
\ee
where
\be
C(t,s)=\langle \sigma _i(t)\sigma _i(s)\rangle
\ee
is the autocorrelation function,
\be
E(t,s)=\langle \sigma_i(t)B_i(s)\rangle,
\ee
and
\be
B_i=-\sum _{[\sigma'']} (\sigma_i-\sigma''_j) w^0([\sigma]\to [\sigma'']).
\label{bj}
\ee
For the dynamic susceptibility one has
\be
T \chi (t,s)=\frac {1}{2} [ C(t,t)-C(t,s)]-
\frac {1}{2}\int _s ^t E(t,t')dt',
\label{nuovachi}
\ee

It is interesting to observe that Eq.~(\ref{nuova}) is completely analogous to
Eqs.~(\ref{primi}) and~(\ref{ax}).
In fact, it can be easily shown that
\be
\frac{d\langle \sigma_i(t)\rangle}{dt}
= \langle B_i(t) \rangle,
\label{ala}
\ee
and that
\be
\frac {\partial C(t,s)}{\partial t} -
\langle B_i(t)\sigma_i(s) \rangle=0.
\label{nuova2}
\ee
Subtracting this from Eq.~(\ref{nuova}) we finally arrive at Eq.~(\ref{primi})
where $A(t,s)$ is given by
\be
A(t,s)=\frac{1}{2}\left [\langle \sigma_i(t)B_i(s)\rangle-\langle B_i(t)\sigma_i(s) \rangle \right ].
\label{asym}
\ee
Eqs.~(\ref{primi}) and~(\ref{asym}) are the main result of this Section. They are 
identical to Eqs.~(\ref{primi}) and~(\ref{ax}) for Langevin dynamics, since the observable $B$ entering in the
asymmetries~(\ref{ax}) and~(\ref{asym}) plays the same role in the two cases. In fact, Eq.~(\ref{ala})
is the analogous of 
\be
{\partial \langle \phi(\vec{x},t)\rangle \over \partial t} = 
\langle B\left [\phi(\vec{x},t)\right ]\rangle
\label{bb4}
\ee
obtained from Eq.~(\ref{I1}) after averaging over the noise.

In summary, Eq.~(\ref{primi}) is a relation between the 
response function and correlation functions of the unperturbed
kinetics, which generalizes the FDT. 
Eq.~(\ref{primi}) applies to a wide class of systems:
Besides being obeyed by soft and hard spins,
it holds both for COP and NCOP
dynamics. Moreover, as it is clear by its derivation,
it does not require any 
particular assumption 
on the Hamiltonian nor on the form of the unperturbed transition rates,
and can be easily generalized~\cite{nat} to intrinsically non-equilibrium systems
where the transition rates do not obey detailed balance.
Finally, let us briefly discuss (for details see Ref.~\cite{noialg})
the differences between the results discussed
insofar and those obtained  
by Chatelain~\cite{chat}, Ricci-Tersenghi~\cite{ricci}, 
Diezemann~\cite{diez} and Crisanti and Ritort~\cite{Crisanti2002}.
Also in these papers, in fact, the response function is related 
to unperturbed correlation functions but, differently from
those appairing in Eqs.~(\ref{nuova},\ref{nuovachi}),
these functions must be computed on a system which
evolves with an {\it ad hoc} kinetic rule, different from
that of the original unperturbed system, which is introduced with the
sole purpose of evaluating the response function. It can be shown that
this corresponds, in the averaging procedure, to consider
only a subset of trajectories of the original unperturbed system.
Therefore, although the results of Refs.~\cite{chat,ricci,diez,Crisanti2002}
are important, both for computational and
analytical calculations, 
they cannot be regarded as generalizations of the FDT 
in the sense of Eq.~(\ref{primi}) because
the response function is not related to correlation functions of the
unperturbed system. 

\section{Fluctuation dissipation relation} \label{flucdisrel}

In the previous Section we have shown that in the cases considered
the integrated response function out of equilibrium is not only related
to the autocorrelation function but also to the correlation
$E(t,s)$ by means of Eq.~(\ref{nuovachi}).
A very useful tool for the
study of slow relaxation phenomena has been introduced by
Cugliandolo and Kurchan~\cite{Cugliandolo93} through the off-
equilibrium fluctuation dissipation relation (FDR). This was
introduced as a direct relation between $\chi (t,s)$ and $C(t,s)$ as follows: 
Given that $C(t,s)$ is a
monotonously decreasing function of $t$, for fixed $s$ it is
possible to invert it and write
\be
\chi(t,s)=\widetilde{\chi}(C(t,s),s).
\ee

Then, if for a fixed value of $C(t,s)$ there exists the
limit
\be
\lim_{s\rightarrow\infty}\widetilde{\chi}(C,s)=S(C)\label{1.20}
\ee
the function $S(C)$ gives the fluctuation dissipation
relation. In the particular case of equilibrium dynamics, FDT is
recovered and $S(C)=[C(\tau=0)-C]/T$. Originally introduced in the
study of the low temperature phase of spin glass mean-field
models, the fluctuation dissipation relation has been found in
many other instances of slow relaxation~\cite{Crisanti2002}.

\section{Statics from dynamics} \label{statdingen}

One of the main reasons of interest in the fluctuation dissipation
relation is that it may provide a link between dynamic and static
properties, and in particular with the equilibrium overlap probability
function 
\be
P(q)=\frac{1}{Z^2}\sum _{[\sigma],\sigma ']}\exp \left \{
-\frac{1}{T}\left [H([\sigma ])+H([\sigma '])\right ]\right \}
\delta \left (Q([\sigma ],[\sigma '])-q\right )
\ee
where $Z$ is the partition function and 
$Q([\sigma ],[\sigma '])=1/N \sum _i \sigma _i \sigma _i'$
is the overlap between two configurations $[\sigma ]$ and $[\sigma ']$.
For slowly relaxing systems this is
established in general by a theorem by Franz {\it et al.}~\cite{Franz98} 
stating that

\begin{enumerate}

\item if $S(C)$ exists

\item if $\lim_{t \rightarrow \infty} \chi(t,s) = \chi_{eq}$,
         $\chi _{eq}$ being the equilibrium susceptibility

\end{enumerate}

then the off-equilibrium fluctuation dissipation relation can be connected
to equilibrium properties through  
\be
\left.-T\frac{d^2 S(C)}{d C^2}\right|_{C=q}=\widetilde{P}(q),
\label{4.3}
\ee
where $\widetilde{P}(q)$ is the overlap probability function
in the equilibrium state obtained in the limit in which the
perturbation responsible of $\chi(t,s)$ is made to vanish. 
The relation between $\widetilde{P}(q)$ and the unperturbed
overlap function $P(q)$ must be considered carefully. 
This implies the notion of stocastic stability~\cite{Guerra}.
In a stochastically stable system the equilibrium state
in the presence of a perturbation, in the limit of a
vanishing perturbation, is the same as that of the
corresponding unperturbed system.
Notice that, while stochastic stability
is always expected for ergodic systems, this property is
far from being trivial when more ergodic components
are present, as it is easily understood by considering
the Ising model perturbed by an external magnetic
field.
If a system is
stochastically stable then $P(q) =\widetilde{P}(q)$. 
A milder statement of
stochastic stability is that $\widetilde{P}(q)$ coincides with
$P(q)$ up to the effects of a global symmetry which might be
removed by the perturbation. For instance, in  the Ising case, where the
perturbation breaks the up-down symmetry, defining 
\be
\widehat{P}(q)= 2\theta(q)P(q) 
\label{4.00} 
\ee 
the system is stochastically stable in the sense that $\widetilde{P}(q) =
\widehat{P}(q)$. In conclusion, if the system is stochastically stable
Eq.~(\ref{4.3}) holds with $\widehat{P}(q)$ on the right hand side,
establishing a connection between the FDR and the equilibrium
properties of the unperturbed state.
On the other hand, if the system is not stochastically stable,
$\widetilde{P}(q)$ is not related neither to $P(q)$ nor to
$\widehat{P}(q)$. As we shall see in Sec.~\ref{failstoc}, this is the case of the mean
spherical model. 

\begin{figure}[htpb]
\begin{center}
   \includegraphics[width=12cm]{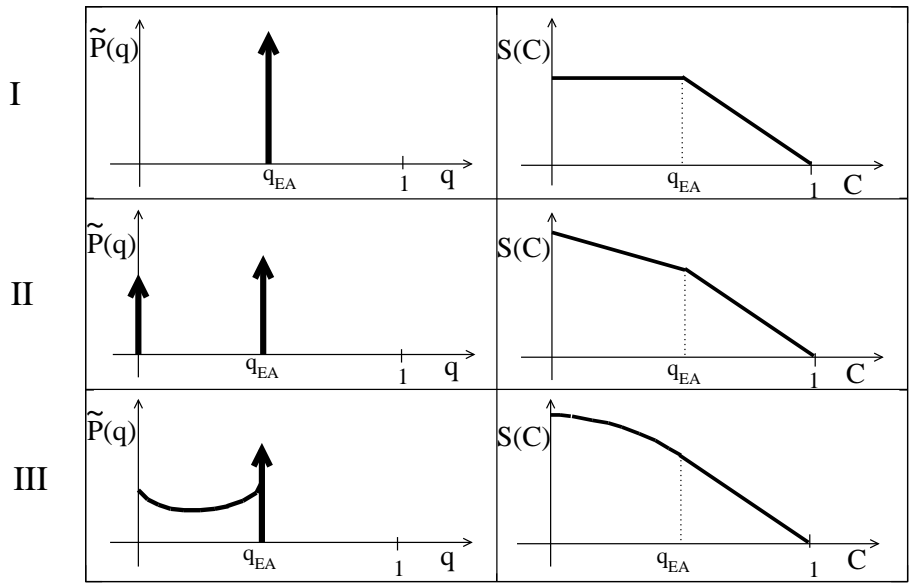}
\vspace{1.5cm}
   \caption{Classification of systems according to their $\widetilde P(q)$
           (left) or, equivalently, on $S(C)$ (right), following ref.~\cite{Ricci99}.
           Bold arrows represent $\delta $ functions. Rows denoted as I, II, and III
           describe the three classes of systems discussed in the text.}
\label{figclass}
\end{center}
\end{figure}

With this link between statics and dynamics one can translate~\cite{Ricci99}
to the dynamics the usual classification of complex systems based on the 
kind of replica symmetry breaking~\cite{Mezard87}. According to this categorization
a first class of systems
are those whose low temperature phase is characterized by two pure state which
are related by a global spin inversion. As will be discussed in 
Sec.~\ref{statdinferro}, these systems without
replica symmetry breaking are described by a $\widetilde{P}(q)$ with
a single $\delta$-function centered on the Edwards-Anderson order parameter
$q_{EA}$ (the squared magnetization, in ferromagnetic systems),
and their FDR, according to Eq.~(\ref{4.3}) is a broken line with an
horizontal part. This situation is shown in Fig.~\ref{figclass}, upper part (I).
A second class of system are those where a transition with a single
step of replica symmetry breaking occurs, as $p$-spins with $p>2$
in mean field, binary mixtures of soft spheres~\cite{Parisi99}
or Lennard-Jones mixtures~\cite{Barrat99}. In these systems
$\widetilde{P}(q)$ is made of two $\delta$-functions, one centered in
the origin and the other around a finite $q_{EA}$. Their FDR is
made of two straight lines with finite slopes, as
shown in Fig.~\ref{figclass} in the central panel (II). 
Systems as the Edwards-Anderson model in mean field fall into a third
category, for which $\widetilde{P}(q)$ is different from zero in a whole
range $q\in [0,q_{EA}]$ with a delta function on $q_{EA}$. 
These systems have a FDR with a straight line and a bending curve,
as shown in Fig.~\ref{figclass}, lower part (III). 

\section{Phase ordering} \label{phord}

Phase ordering~\cite{Bray94}
is usually regarded as the simplest instance of slow relaxation,
where concepts like scaling and aging, which are the hallmarks of glassy
behavior~\cite{Cugliandolo2002}, can be more easily investigated. 
However, next to the similarities
there are also fundamental differences~\cite{CCY} which require to keep phase
ordering well distinct from the out of equilibrium behavior in
glassy systems, both disordered and non disordered. The main
source of the differences is the simplicity of the free energy
landscape in the case of phase ordering compared to the complexity
underlying glassy behavior. 

Besides the obvious motivation that the basic,
paradigmatic cases need to be thoroughly understood, an
additional reason for studying phase ordering, 
among others, is that in some cases the
existence of complex slow relaxation is identified through the
exclusion of coarsening. 
An example comes from the long standing controversy
about the nature of the low temperature phase of finite
dimensional spin glasses. One argument in favor of replica
symmetry breaking is that the observed behavior of the response
function is incompatible with coarsening~\cite{Franz98,Ricci99}.
This might well be the case; however, for the argument to be
sound, the understanding of the out of equilibrium behavior of the
response function during phase ordering needs to be up to the
level that such a delicate issue demands.

In this Section we present an overview of the
accurate investigation of the response function in phase ordering
that we have carried out in the last few years. Focusing on the
integrated response function~(\ref{integ}), 
or zero field cooled magnetization (ZFC) in the language of magnetic systems, 
it will be argued that the response
function in phase ordering systems is not as trivial as it is
believed to be and, after all, it is not the quantity best suited
to highlight the differences between systems with and without
replica symmetry breaking. In fact, as discussed 
in Secs.~\ref{statdinferro},\ref{failstoc},
there are cases in which phase ordering, 
and therefore a replica symmetric low temperature state,
are compatible with a non trivial ZFC. When this happens there is
no connection between static and dynamic properties. Phase
ordering systems offer examples of two distinct mechanism for the
lack of this important feature of slow relaxing systems, 
stochastic instability and the
vanishing of the scaling exponent of ZFC. 

Let us first briefly recall the main features of a phase ordering
process. Consider a system, like a ferromagnet, with order
parameter (vector or scalar, continuous or discrete) $\phi(\vec x)$
and Hamiltonian ${\cal H}[\phi(\vec x)]$ such that below the
critical temperature $T_C$ the structure of the equilibrium state
is simple. For example, in the scalar case, there are two pure
ordered states connected by inversion symmetry. The form of the
Hamiltonian can be taken the simplest compatible with such a
structure, like Ginzburg-Landau-Wilson for continuous spins
or the nearest neighbors Ising Hamiltonian for discrete spins.

Let us generalize the definition~(\ref{prii}) to the space and time
dependent correlation function

\be
C(\vec{r},t,s)=\langle\phi(\vec x,t)\phi(\vec x',s)\rangle-
\langle\phi(\vec x,t)\rangle\langle\phi(\vec
x',s)\rangle\label{1.1}
\ee
where the average is taken over initial condition and thermal
noise, and $\vec{r}= \vec{x} - \vec{x'}$. 
We use the notation $C(\vec r=0,t,s)=C(t,s)$,
and similarly for the response functions defined below.  
The linear response function conjugated to $C(\vec r,t,s)$ is
given by

\be
R(\vec{r},t,s)=\left.\frac{\delta\langle\phi(\vec{x},t)\rangle}{\delta h(\vec{x'},s)}\right|_{h=0},
\label{1.2}
\ee
where $h(\vec x,t)$ is a space-time dependent external magnetic field
and the integrated response function is defined by

\be
\chi(\vec{r},t,s)= \int_{s}^t ds R(\vec{r},t,s).\label{1.3}
\ee

\subsection{Dynamics over phase space: equilibration versus
falling out of equilibrium} \label{falling}

For a temperature $T$ below $T_C$, in the thermodynamic limit, the
phase space $\Omega=\{[\phi(\vec x)]\}$ may be regarded as the
union of three ergodic components~\cite{Palmer} $\Omega=\Omega_+
\cup \Omega_- \cup \Omega_0$, where $\Omega_{\pm}$ and $\Omega_0$
are the subsets of configurations with magnetization $\lim_{V
\rightarrow \infty} \frac{1}{V}\int_V d \vec{x}\phi(\vec x)$
positive, negative and vanishing, respectively. Denoting by
$\rho_{\pm}[\phi(\vec x)]$ the two broken symmetry pure states,  
whose typical configurations are schematically
represented in Fig.~\ref{brokensymmetry},
all equilibrium states are the convex linear combinations of
$\rho_{\pm}$. In particular, the Gibbs state is the symmetric
mixture $\rho_G[\phi(\vec x)]=\frac{1}{Z}\exp (-{\cal H}[\phi(\vec
x)]/T)= \frac{1}{2}\rho_+[\phi(\vec x)] +
\frac{1}{2}\rho_-[\phi(\vec x)]$. The $\Omega_{\pm}$ components
are the domains of attraction of the pure states with
$\rho_+(\Omega_+)= \rho_-(\Omega_-)= 1$ and $\Omega_0$ is the
border in between them, with zero measure in any of the
equilibrium states.

When ergodicity is broken, quite different behaviors may
arise~\cite{Palmer}  depending on the initial condition
$\rho_0[\phi(\vec x)]= \rho([\phi(\vec x)],t=0)$. Here, we
consider the three cases relevant for what follows, assuming that
there are not explicit symmetry breaking terms in the equation of
motion:

\begin{enumerate}

\item {\it equilibration to a pure state}

if $\rho_0(\Omega_{+})=1$ or $\rho_0(\Omega_{-})=1$, in the time evolution
configurations are sampled from either one of $\Omega_{\pm}$ and
$\rho([\phi(\vec x)],t)$ equilibrates to the time independent pure
state $\rho_{\pm}[\phi(\vec x)]$ within the finite relaxation time
$t_{eq} \sim \xi^z$, where $\xi$ is the equilibrium correlation
length and $z$ is the dynamic exponent. The correlation function is the same
in the two ergodic components and, after equilibration, is time
translation invariant

\be
   C_{st}(\vec{r},\tau )=\langle\phi(\vec x,t)\phi(\vec x',s)\rangle_{\pm}-M^2\label{1.4}
\ee

where $\langle \phi(\vec x)\rangle _{\pm} = \pm M$ is the
spontaneous magnetization. For large distances $r \gg \xi$ and
time separations $t-s \gg t_{eq}$, the clustering property
$\langle \phi(\vec x,t)\phi(\vec x',s)\rangle _{\pm}$ $=\langle
\phi(\vec x,t) \rangle _{\pm} \langle\phi(\vec x',s)\rangle
_{\pm}$ is obeyed and the correlations decay to zero, as required
by ergodicity (see Fig.~\ref{scaltemp}).

\vspace{3cm}
\begin{figure}[htpb]
\begin{center}
   \includegraphics[width=8cm]{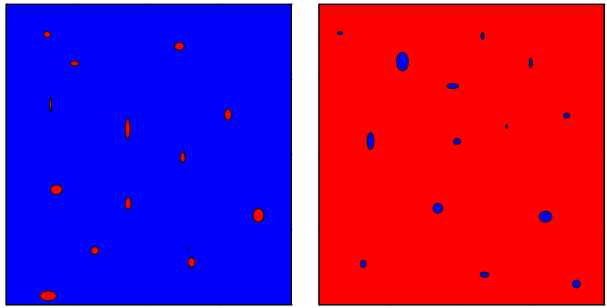}
   \vspace{1cm}
   \caption{Typical configurations of a binary system after
            equilibration to the pure states $\rho_{+}[\phi(\vec x)]$
            or $\rho_{-}[\phi(\vec x)]$ (left and right panel).}
\label{brokensymmetry}
\end{center}
\end{figure}

\item {\it equilibration to the Gibbs state}

if $\rho_0(\Omega_+)=\rho_0(\Omega_-) =1/2$, then configurations
are sampled evenly from both disjoint components  $\Omega_+$ and
$\Omega_-$. The probability density $\rho([\phi(\vec x)],t)$
equilibrates now to the Gibbs state $\rho_G[\phi(\vec x)]$ with
the same relaxation time $t_{eq}$ as in the relaxation to the pure
states. Broken ergodicity shows up in the large distance and in
the large time properties of the correlation function. After
equilibration, one has

\be
   C_G(\vec{r},\tau )=C_{st}(\vec{r},\tau )+M^2\label{1.9}
\ee

from which follows that correlations do not vanish asymptotically
or that the clustering property is not obeyed

\be
   \lim_{r\rightarrow\infty}C_G(\vec{r},\tau )=
\lim_{\tau \rightarrow\infty}C_G(\vec{r},\tau)=M^2.\label{1.10}
\ee

\item {\it falling out of equilibrium over the border~\cite{Laloux,Newman}}

If $\rho_0(\Omega_0)=1$, for the infinite system $\rho(\Omega_0,t)
=1$ also at any finite time after the quench. Namely, the system
does not equilibrate since in any equilibrium state the measure of
$\Omega_0$ vanishes. Phase ordering corresponds to this case. In
fact, the system is initially prepared in equilibrium at very high
temperature (for simplicity $T_I=\infty$) and at the time $t=0$ is
suddenly quenched to a final temperature $T$ below $T_C$. In the
initial state the probability measure over phase space is uniform
$\rho_0 [\phi(\vec x)] = 1/ | \Omega |$, implying that the initial
configuration at $t=0$ belongs almost certainly to $\Omega_0$,
since with a flat measure $| \Omega_0 |$ is overwhelmingly larger
than $| \Omega_{\pm}|$.

The morphology of typical configurations visited as the system
moves over $\Omega_0$ is a patchwork of domains of the two
competing equilibrium phases, which coarsen as the time goes on,
as schematically shown in Fig.~\ref{domains}.
The typical size of domains grows with the power law $L(t) \sim
t^{1/z}$, where $z=2$ (independent of dimensionality) for dynamics
with non conserved order parameter~\cite{Bray94}, as it will be
considered here. The sampling of configurations of this type is
responsible of the peculiar features of phase ordering. At a given
time $s$ there remains defined a length $L(s)$ such that for
space separations $r \ll L(s)$ or for time separations $t-s
\ll s$ intra-domains properties are probed. Then, everything
goes as in the case 2 of the equilibration to the Gibbs state,
ergodicity looks broken and the correlation function obeys
Eq.~(\ref{1.9}). Conversely, for  $r \gg L(s)$ or  $t/s \gg
1$, inter-domains properties are probed, ergodicity is restored
(as it should be, since evolution takes place within the single
ergodic component  $\Omega_0$) and eventually the correlation
function decays to zero. However, the peculiarity is that if the
limit $s \rightarrow \infty$ is taken before $r \rightarrow
\infty$, in the space sector ergodicity remains broken giving
rise, for instance, to the growth of the Bragg peak in the equal
time structure factor.

\begin{figure}[htpb]
\begin{center}
   \includegraphics[width=8cm]{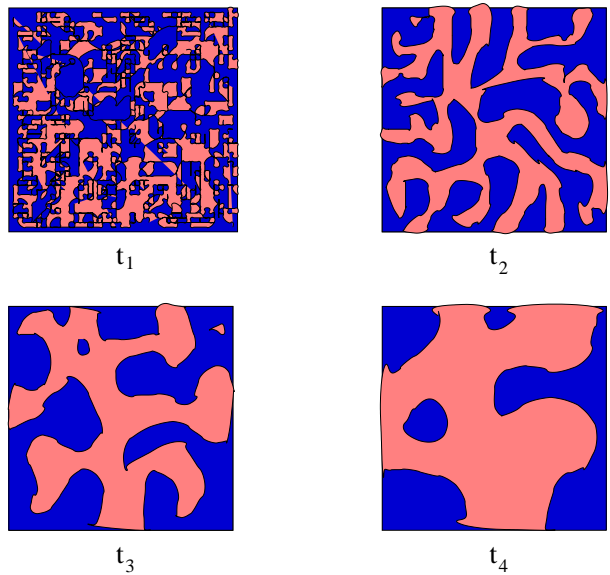}
   \caption{Configurations of a coarsening system at different
           times $t_1<t_2<t_3<t_4$.}\label{domains}
\end{center}
\end{figure}

According to this picture, the correlation function can be written
as the sum of two contributions

\be
   C(\vec{r},t,s)=C_{st}(\vec{r},\tau )+C_{ag}(\vec{r},t,s)\label{1.131}
\ee

where the first one is the stationary contribution of
Eq.~(\ref{1.4}) describing equilibrium fluctuations in the pure
states and the second one contains all the out of equilibrium
information. The latter one is the correlation function of
interest in the theory of phase ordering where, in order to
isolate it, zero temperature quenches are usually considered as a
device to eliminate the stationary component. It is now well
established that $C_{ag}(\vec{r},t,s)$ obeys scaling in the
form~\cite{Furukawa}

\be
   C_{ag}(\vec{r},t,s)=\widehat{C}(r/L(s),t/s)\label{1.132}
\ee

with $\widehat{C}(x,y)=M^2$ for $x<1$ and $y \sim 1$, while

\be
   \widehat{C}(r/L(s),t/s)\sim(t/s)^{-\lambda/z}h(r/L(s))
\ee

for large time separation~\cite{Bray94}, where $\lambda$ is the
Fisher--Huse exponent.

\end{enumerate}

\subsection{Zero field cooled magnetization} \label{zerofc}

Let us next consider what happens when a time independent external
field $h(\vec{x},s)$ is switched on at the time $s$. To linear
order the expectation value of the order parameter at the later time $t$
is given by

\be
   \langle\phi(\vec{x},t)\rangle_{h}=\langle\phi(\vec{x},t)\rangle_{0}+
\int d\vec{x'}\int ds R(\vec{x}-\vec{x'},t,s)h(\vec{x'},s)
\label{2.2}
\ee
If $h(\vec{x},s)=h(\vec x)\theta (t-s)$ is a random field switched on and kept constant
from $s$ onwards, with expectations
\be
\overline{h(\vec{x})}=0
\label{exp1}
\ee
\be
\overline{h(\vec{x})h(\vec{x'})}=
h^2\delta(\vec{x}-\vec{x'})
\label{exp2}
\ee
then one has
\be
   \chi(\vec{x}-\vec{y},t,s)=h^{-2}\overline{\langle\phi(\vec{x},t)\rangle_{h}h(\vec{y})}.
\label{2.3}
\ee
Namely, ZFC is the correlation at the time $t$ of the
order parameter with the random external field.

Going to the three processes considered above, and restricting attention from now on, for
simplicity, to the case of coincident points ($\vec{r} = 0$) 

\begin{enumerate}
\item after equilibration in the pure state has occurred and the
stationary regime has been entered, the order parameter correlates
with the external field via the equilibrium thermal fluctuations, FDT is obeyed

\be
   \chi_{st}(\tau)=\frac{1}{T}\left[C_{st}(\tau=0)-C_{st}(\tau)\right]\label{2.4}
\ee

and since $C_{st}(\tau)$ decays to zero for $\tau >
t_{eq}$, over the same time scale $\chi_{st}(\tau)$
saturates to

\be
   \lim_{t\rightarrow\infty}\chi_{st}(\tau )=\chi_{eq}=\frac{1}{T}C_{eq}
\label{2.5}
\ee
which is the susceptibility computed in the final equilibrium
state $\rho_{\pm}[\phi(\vec{x})]$ (see Fig.~\ref{scaltemp}).

\item As far as ZFC is concerned, there is no difference between
the relaxation to the mixed Gibbs state and the relaxation to a
pure state. Hence, FDT is satisfied and can be written both in
terms of $C_{st}$ or $C_G$ since, as Eq.~(\ref{1.9}) shows, they
differ by a constant.

\item In the phase ordering process the system stays out of
equilibrium, so it useful to write ZFC as the sum of two
contributions~\cite{Bouchaud97}

\be
   \chi(t,s)=\chi_{st}(\tau)+\chi_{ag}(t,s)
\label{2.7}
\ee
where $\chi_{st}(\tau)$ satisfies  Eq.~(\ref{2.4}) and
$\chi_{ag}(t,s)$ represents the additional out of
equilibrium response. In connection with this latter contribution
there are two basic questions

i) how does it behave with time

ii) what is the relation between  $\chi_{ag}$ and  $C_{ag}$, if any.

\end{enumerate}

\subsubsection{Scaling hypothesis} \label{scalh}

Since ZFC measures the growth of correlation between the order
parameter and the external field, the first question raised above
addresses the problem of an out of equilibrium mechanism for this
correlation, in addition to the thermal fluctuations accounting
for $\chi_{st}$. 
The starting point for
the answer is the assumption of a scaling form

\be
   \chi_{ag}(t,s)\sim s^{-a_{\chi}}\widehat{\chi}_{ag}(t/s)\label{2.800}
\ee
which is the counterpart of Eq.~(\ref{1.132}) for the
correlation function.

\indent The next step is to make statements on the exponent
$a_{\chi}$ and on the scaling function $\widehat{\chi}_{ag}(x)$  .
There exists in the literature an estimate of $a_{\chi}$ based on
simple reasoning. What makes phase ordering different from
relaxation in the pure or in the Gibbs state is the existence of
defects. The simplest assumption is that $\chi_{ag}(t,s)$ is
proportional to the density of
defects~\cite{Barrat98,Franz98,Ricci99}. This implies

\be
   a_{\chi}=\delta\label{2.82}
\ee

\noindent where the exponent $\delta$ regulates the time
dependence of the density of defects $\rho (t)  \sim
L(t)^{-n} \sim t^{-\delta}$, namely

\be
   \delta=n/z\label{3.2}
\ee

\noindent with $n=1$ for scalar and $n=2$ for vector order
parameter~\cite{Bray94}.

According to this argument $a_{\chi}$ should be independent of
dimensionality. This conclusion is not corroborated by the
available exact, approximate and numerical results. On the basis
of exact analytical solutions for the $d=1$ Ising
model~\cite{Lippiello2000,Godreche} and for the large $N$
model~\cite{Corberi2002}, approximate analytical results based on
the Gaussian auxiliary field (GAF)
approximation~\cite{Berthier99,Corberi2001} and numerical results
from simulations~\cite{Corberi2001,prl,preprint,generic,clock} with $d=2,3,4$, 
we have argued that

\be %\label{3.8}
    a_{\chi} = \left \{ \begin{array}{ll}
        \delta \left( \frac{d-d_L}{d_U-d_L} \right ) \qquad $for$ \qquad d < d_U \\
        \delta  \qquad $with log corrections for$ \qquad d = d_U \\
        \delta  \qquad $for$ \qquad  d > d_U
        \end{array}
        \right .
        \label{3.1}
\ee

\noindent where $d_L$ and $d_U > d_L$ do depend on the system in
the following way

\begin{itemize}

\item $d_L$ is the dimensionality where  $a_{\chi}=0$.
In the Ising model $d_L=1$, while in the large
$N$ model $d_L =2$. The speculation is that in general $d_L=1$ for
systems with discrete symmetry and $d_L =2$ for systems with
continuous symmetry, therefore suggesting that $d_L$ coincides with the lower
critical dimensionality of equilibrium critical phenomena,
although the reasons for this identification are far from clear.

\item $d_U$ is a value of the dimensionality specific of ZFC and
separating $d <d_U$, where $a_{\chi}$ depends on $d$, from $d >d_U$
where $a_{\chi}$ is independent of dimensionality and
Eq.~(\ref{2.82}) holds. The existence of $d_U$ is
due~\cite{preprint} to a mechanism, i.e. the existence of a
dangerous irrelevant variable, quite similar (including
logarithmic corrections) to the one leading to the breaking of
hyperscaling above the upper critical dimensionality in static
critical phenomena. However, $d_U$ cannot be identified with the
upper critical dimensionality since we have found, so far,  $d_U
=3$ in the Ising model and  $d_U =4$ in the large $N$ model. 
In the scalar case it may be argued~\cite{generic,henkel04} that $d_U$ coincides with the 
dimensionality $d_R=3$ such that interfaces do roughen for
$d \leq d_R$ and do not for $d > d_R$. This will be discussed
in Sec.~\ref{roughening}. 
A general criterion for establishing the value of $d_U$, however, is not yet
known.

\end{itemize}

\begin{figure}[htpb]
\begin{center}
   \includegraphics[width=8cm]{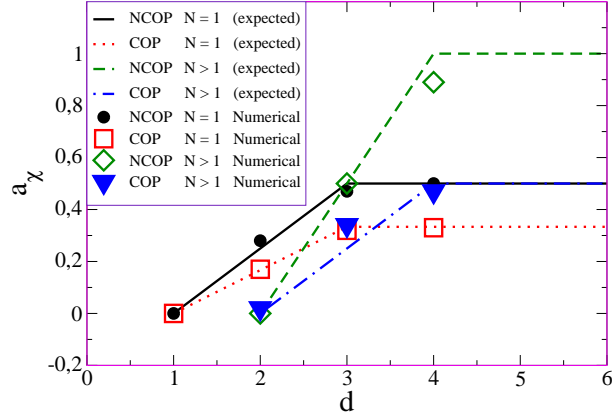}
   \caption{Exponent $a_{\chi}$ in different coarsening
            systems with scalar and vector order parameter,
            non-conserved and conserved order parameter~\cite{Bray94}.
           The continuous lines represent Eq.~(\ref{3.1}),
	   while the dots are the values from
	   simulations\cite{preprint,generic}.}
\label{anumerico}
\end{center}
\end{figure}

The validity of Eq.~(\ref{2.800}) with $a_{\chi}$ given by
Eq.~(\ref{3.1}) has been checked, in addition to the cases where
analytical results are available, with very good accuracy in the
simulations of the Ising and clock model
and of the time dependent
Ginzburg-Landau equation~\cite{Corberi2001,prl,preprint,generic,clock}. 
The values of
$\delta$, $d_L$ and $d_U$ obtained for the different systems are
collected in Table \ref{tabella} and the behavior of $a_{\chi}$ as
dimensionality is varied is displayed in Fig.\ref{anumerico}.

\begin{table}[th]      %Table~1.1
%\tbl{Parameters entering Eq.(\ref{3.1}) in various models.}
\begin{tabular}{|l|c|c|c|}
    \hline           & Ising & GAF & $N=\infty$ \\
    \hline  $\delta$ & 1/2   & 1/2 &  1 \\
    \hline  $d_L$    &  1    &  1  &  2 \\
    \hline  $d_\chi$ &  3    &  2  &  4 \\
    \hline
\end{tabular}
\caption{Parameters entering Eq.(\ref{3.1}) in various models.}
\label{tabella}
\end{table}

\vspace{1cm}

\subsubsection{Roughening of interfaces} \label{roughening}

Apart from the few exact solutions mentioned above
there is not a general derivation of Eq.~(\ref{3.1}) which,
at this stage remains a phenomenological formula.
For the case of a scalar order parameter,
an argument has been proposed~\cite{generic,henkel04} explaining the dependence of
$a_{\chi}$ on $d$ in terms of the roughening properties
of the interfaces.
It is based on two simple physical ingredients:
{\bf a)} the aging response is given by the
density of defects $\rho(t)$ times the response of a single
defect~\cite{Corberi2001} $\chi_{ag}(t,s)=\rho(t)\chi_{ag}^s(t,s)$ and
{\bf b)} each defect responds to the perturbation
by optimizing its position with respect to the external field in a quasi-equilibrium way.
In $d=1$ this occurs via a displacement of the defect~\cite{Corberi2001}.
In higher dimensions, since defects are spatially extended,
the response is produced by a deformation of the defect shape.

\begin{figure}
\includegraphics[angle=0,width=7cm,clip]{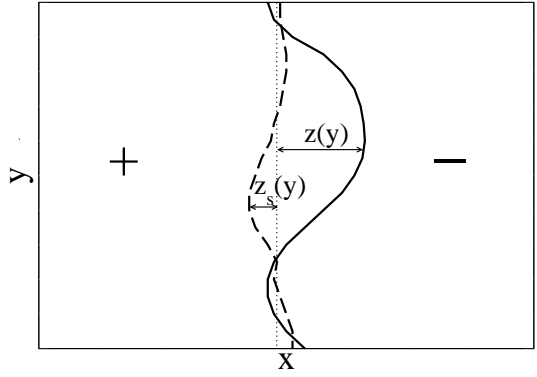}
\caption{Configurations with a single interface at time $s$ (dashed line)
and at time $t$ (continuous line).}
\label{fig_interface}
\end{figure}

We develop the argument for a 2-d system,
the extension to arbitrary $d$ being straightforward.
A defect is a sharp interface separating two domains
of opposite magnetization. In order to analyze $\chi_{ag}^s(t,s)$ we consider
configurations with a single defect as depicted in Fig.~\ref{fig_interface}.
The corresponding integrated response function 
reads~\cite{Corberi2001} 
$\chi_{ag}^s(t,s) =1/(h^2 {\cal L}^{d-1})\int d{\text x} d{\text y} \,
\overline{\langle \phi({\text x},{\text y}) \rangle h({\text x},{\text y})}$, where $\phi({\text x},{\text y})$ 
is the order parameter field which saturates to $\pm 1$ in the bulk of domains,
and ${\text x},{\text y}$ are space coordinates. 
$h({\text x},{\text y})$ is the external random field with
expectations~(\ref{exp1},\ref{exp2}), and 
${\cal L}$ is the linear system size.
The overbar and angular brackets denote averages over the
random field and thermal histories, respectively.
With an interface of shape $z_s({\text y})$ at time $s$ (Fig.~\ref{fig_interface}), 
we can write 
$\chi_{ag}^s(t,s) = - 1/(h^2 {\cal L}^{d-1})
\overline{\int_{\{z \}} E_h \, P_h(\{z({\text y})\},t)}$,
where $P_h(\{z({\text y})\},t)$ is the probability that an interface profile $\{z({\text y})\}$
occurs at time $t$ and
$E_h = -\int _0 ^{\cal L} d{\text y} \int_{z_s({\text y})}^{z({\text y})} d{\text x} h({\text x},{\text y}) 
\mbox{sign} [z({\text y})-z_s({\text y})]$
is the magnetic energy. 
We now introduce assumption b) making the ansatz for the correction to the
unperturbed probability $P_0(\{z({\text y}) \},t)$ in the form of a Boltzmann factor
$P_h(\{z({\text y})\},t)=P_0(\{z({\text y})\},t) \exp(-E_h/T) \simeq P_0(\{z({\text y})\},t)
[1-E_h/T]$. Then
$\chi_{ag}^s(t,s)=-1/(h^2 {\cal L}^{d-1})\overline {\int_{\{z \}} E_h(1- E_h/T)
P_0(\{z({\text y})\},t)}$.
Taking into account that the term linear in $E_h$ vanishes by symmetry and 
neglecting $z_s({\text y})$ with respect to $z({\text y})$ for $t\gg s$, we eventually find 
$T\chi_{ag}^s(t,s)=
{\cal L}^{1-d} \int_{\{z \}} \int _0^{\cal L} d{\text y}
\vert z({\text y})\vert P_0(\{z({\text y})\},t)$. 
This defines a length which scales as
the roughness of the interface given by
$W(t) = [{\cal L}^{1-d} \int_{\{z \}} \int d{\text y} z({\text y})^2 
P_0(\{z({\text y})\},t)]^{1/2}$.
The behavior of $W(t)$ in the coarsening process can be inferred
from an argument due to Villain~\cite{Abraham89}.
In the case $d\le 3$, when interfaces are
rough~\cite{Rough}, for NCOP one has $W(t) \sim t^{(3-d)/4}$, while for
COP $W(t) \sim t^{(3-d)/6}$, with logarithmic corrections in both cases 
for $d=3$. 
For $d>3$ interfaces are flat and $W(t)\simeq const.$
Finally, multiplying $\chi_{ag}^s$ by 
$\rho (t) \sim L(t)^{-1}$ 
Eq.~(\ref{3.1}) is recovered~\cite{note2} and 
$d_U$ is identified with the roughening dimensionality $d_R=3$.

\subsection{Statics from dynamics} \label{statdinferro}

We may now check if, and how, the connection between statics
and dynamics discussed in Sec.~\ref{statdingen} is realized in phase
ordering systems. In
the following we shall consider $d \geq d_L$.

In order to search for $S(C)$ in the case of phase ordering,
let us set $\vec{r}=0$ in Eq.~(\ref{1.132}) and let us eliminate
$t/s$ between $\widehat{\chi}_{ag}$ and $C_{ag}$ obtaining

\be
   \chi_{ag}(t,s)\sim s^{-a_{\chi}}\widetilde{\chi}_{ag}(C_{ag}).\label{2.8}
\ee

Then, from Eqs. ~(\ref{2.7},\ref{2.4},\ref{2.8}) one can write the
general relation

\be
   \chi(t,s)=\frac{1}{T}\left[C_{st}(\tau=0)-C_{st}(\tau )\right]+
s^{-a_{\chi}}\widetilde{\chi}_{ag}(C_{ag}).
\ee

\begin{figure}
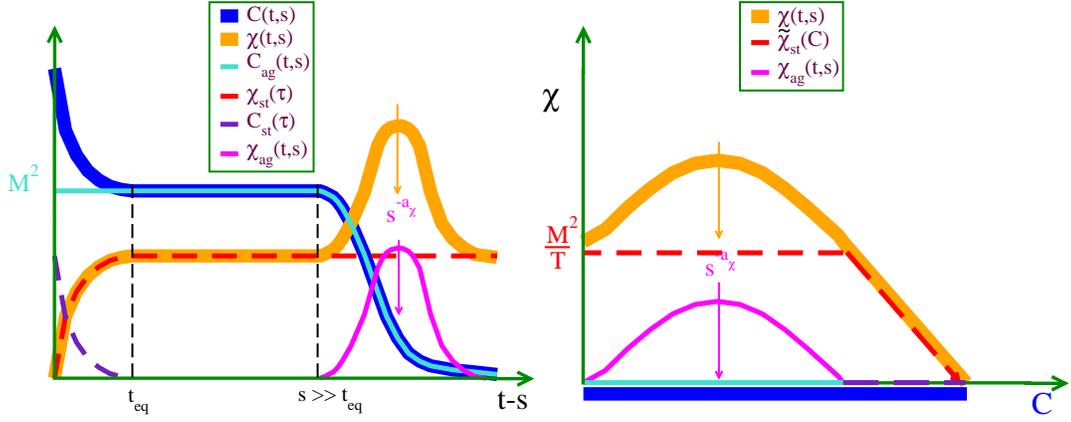

\includegraphics[angle=0,width=7cm,clip]{scaletempi.eps}
\includegraphics[angle=0,width=7cm,clip]{scaletempi2.eps}
\caption{Schematic plot of the behavior of the two time 
functions in the coarsening stage as $t-s$ is varied
keeping $s$ fixed (upper figure), and of the
resulting fluctuation dissipation plot (lower figure).
The stationary parts $C_{st}$, $\chi _{st}$, as discussed 
in Secs.~\ref{falling},\ref{zerofc}, saturate to their
final value on times $\tau\simeq t_{eq}$; the aging parts
 $C_{ag}$, $\chi _{ag}$, according to the scaling forms~(\ref{1.132},\ref{2.800})  
remain constant up to times
$t-s\simeq s$. The magnitude of $\chi _{ag}$ is proportional
to $s^{-a_\chi}$, and decreases as $s$ is increased (shown 
by an arrow in Figure).}
\label{scaltemp}
\end{figure}

Using the identity $\left [ C_{st}(\tau=0) - C_{st}(\tau)\right ] =
\left [ C_{st}(\tau=0)+ M^2  - C_{st}(\tau) - M^2 \right ]$ and
considering that, as shown schematically in Fig.~\ref{scaltemp}, 
in the time interval where $C_{st}(\tau) \neq
0$, i.e. for short times, one can replace $C_{ag}(t/s)$ with
$M^2$ or equivalently $ C_{st}(\tau) + M^2 = C(t,s)$, the above
equation can be rewritten as

\be
   \chi(t,s)=\widetilde{\chi}_{st}(C)+s^{-a_{\chi}}\widetilde{\chi}_{ag}(C_{ag})\label{2.9}
\ee

where the function $\widetilde{\chi}_{st}(C)$ is defined by

\be
    T\widetilde{\chi}_{st}(C) = \left \{ \begin{array}{ll}
        \left [ C(t,t) - C(t,s) \right ]  \qquad $for$ \qquad  M^2 \leq C \leq C(t,t) \\
        \left [ C(t,t) - M^2 \right ]  \qquad $for$ \qquad C < M^2.
        \end{array}
        \right .
        \label{2.10}
\ee

Therefore, from Eq.~(\ref{2.9}) we have that for
phase ordering systems the fluctuation dissipation relation exists
if $a_{\chi} > 0$ (i.e. for $d > d_L$) and it is given by

\be
   S(C)=\widetilde{\chi}_{st}(C).
\label{1.30}
\ee
Computing the derivative in the left hand side of Eq.~(\ref{4.3})
and using Eqs.~(\ref{1.30}) and~(\ref{2.10}), for $d> d_L$ we find

\be
   \left. -T \frac{d^2 S(C)}{d C^2} \right |_{C=q} = \delta
(q-M^2). \label{4.6}
\ee
Coming to statics, in replica symmetric low temperature states, as for
instance in ferromagnetic systems,  the overlap function is always
trivial and, as anticipated in Sec.~\ref{statdingen}, one has
\be
   P(q) = \frac{1}{2} \left [ \delta (q-M^2) + \delta (q+M^2)
\right ], 
\label{4.4}
\ee
with
\begin{equation}
\widetilde{P}(q) = \widehat{P}(q) = \delta (q-M^2).
\label{4.4bis}
\end{equation}
as shown in Fig.~\ref{figclass} (I) (we recall that $q_{EA}=M^2$
in this case). 
From Eqs.~(\ref{4.4bis},\ref{4.6}), therefore, Eq.~(\ref{4.3}) is satisfied, 
and the connection between statics and dynamics holds.

\begin{figure}[htpb]
\begin{center}
   \includegraphics[width=8cm]{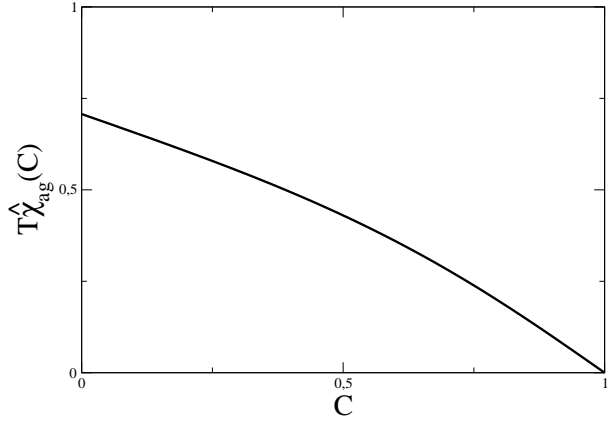}
   \caption{Fluctuation dissipation plot for the 1d-Ising model.}
   \label{ising1d}
\end{center}
\end{figure}

For $a_{\chi} = 0$ a little more care is needed.
Equation~(\ref{2.9}) yields $\chi(t,s) =
\widetilde{\chi}_{st}(C) + \widetilde{\chi}_{ag}(C_{ag})$.
Recalling that $a_{\chi} = 0$ occurs at  $d = d_L$, which
coincides with the lower critical dimensionality, in order to have
a phase ordering process a quench to $T=0$ is required. This, in
turn, implies $C_{st}(t,s) =0$ and $C_{ag}(t,s) =C(t,s)$.
Therefore, using Eq.~(\ref{2.10}) we have
\be
   S(C)=\chi_{eq}^*+\widetilde{\chi}_{ag}(C)\label{1.40}
\ee
where $\chi_{eq}^* = \lim_{T \rightarrow 0}[C(0) -
M^2]/T$ is the $T=0$ equilibrium susceptibility, which vanishes
for hard spins while is different from zero for soft spins.
Therefore the FDR exists also in this case. However,
while for $a_{\chi} >0$ $\chi_{ag}$ eventually disappears and
Eq.~(\ref{2.5}) holds, this is no longer true for $d = d_L$. 
Here $a_{\chi}= 0$ and, consequently, as can be seen from Eq.~(\ref{1.40})
$\chi_{ag}$ gives a contribution to the
response which persists also in the asymptotic time region. 
Then Eq.~(\ref{2.5}), and hence condition (2)
above Eq.~(\ref{4.3} are not fulfilled. Being one of the
hypothesis leading to Eq.~(\ref{4.3}) violated,
the connection between statics and dynamics could not
hold. Actually, in all the model explicitly considered
in the literature~\cite{Corberi2001,preprint,generic}
it turns out that at $d=d_L$
$S(C)$ is a non-trivial dynamical function unrelated
to $\widetilde P(q)$.
For the sake of
definiteness, let us discuss the case of the Ising model with 
$d=1$~\cite{Godreche, Lippiello2000}. In
order to make compatible the two requirements of having an ordered
equilibrium state and a well defined linear response function,
instead of taking the $T \rightarrow 0$ limit it is necessary to
take the limit of an infinite ferromagnetic
coupling~\cite{Lippiello2000}. Then, $P(q)$ and $\widetilde{P}(q)$
are given by Eqs.~(\ref{4.4}) and~(\ref{4.4bis}) with $M^2=1$ at
all temperatures. On the other hand, for any $T$ one also
have~\cite{Lippiello2000} (see Fig.~\ref{ising1d})

\begin{equation}
T\widetilde{\chi}_{ag}(C) = \frac{\sqrt{2}}{\pi} \arctan \left [ \sqrt{2}\cot \left (
\frac{\pi}{2} C \right ) \right ].
\label{4.5}
\end{equation}
This gives

\be
   \left.-T\frac{d^2 S(C)}{d C^2} \right |_{C=q} = \frac{\pi
\cos (\pi q/2)\sin (\pi q/2)}{[2 -\sin (\pi q/2)]^2}.
\label{4.6bis}
\ee
Hence, it is clear that Eq.~(\ref{4.3}) is not verified.
The reason is that the second of the above conditions required for
establishing the connection is not satisfied. In fact, from
Eqs.~(\ref{2.9}) and~(\ref{4.5}), keeping in mind that the limits
$t \rightarrow \infty$ and $C \rightarrow 0$ are equivalent, we
have

\be
   \lim_{t\rightarrow\infty}T\chi(t,s)=1/\sqrt{2}\label{4.8}
\ee

which is responsible of the violation of condition (2) above
Eq.~(\ref{4.3}), since in this case $\chi_{eq}=0$.
Interestingly, a similar behavior is observed~\cite{prlparma} also for
the Ising model on graphs with $T_c=0$, which, in a sense, can
be regarded as being at $d_L$.

\vspace{1cm}

\subsubsection{Role of quenched disorder at $d=d_L$} \label{quenchd}

The behavior of the exponent $a_\chi $, and its vanishing at $d_L=1$  
can be qualitatively interpreted
in terms of the behavior of the response $\chi ^s _{ag}(t,s)$ associated to a single
interface. In $d=1$ it can be shown exactly~\cite{Corberi2001} 
that $\chi ^s _{ag}(t,s)\propto (t-s)^{1/2}$.
Therefore, when computing the total response through 
$\chi _{ag}(t,s)\simeq \rho (t) \chi ^s (t,s)$ the loss of interfaces described by
$\rho (t)$ is exactly balanced by the increase of $\chi ^s (t,s)$, which leads
to a finite $\lim _{t\to \infty }\chi _{ag}(t,s)$. This in turn is responsible
for the breakdown of the condition (2) above Eq.~(\ref{4.3}). 
For $d>d_L$,
instead, the growth of $\chi ^s (t,s)$ is not sufficient~\cite{Corberi2001} to
balance the decrease of $\rho (t)$. This happens because, while in $d=1$ interfaces
are Brownian walkers, free to move in order to
maximize the response, for $d>1$ this issue is contrasted by 
surface tension, restoring the validity of condition (2). 
In this Section how a similar effect, namely
a reduction of the response of interfaces, can be produced, also in $d=1$,
by the presence of quenched disorder. 

\begin{figure}[htpb]
\begin{center}
   \includegraphics[width=8cm]{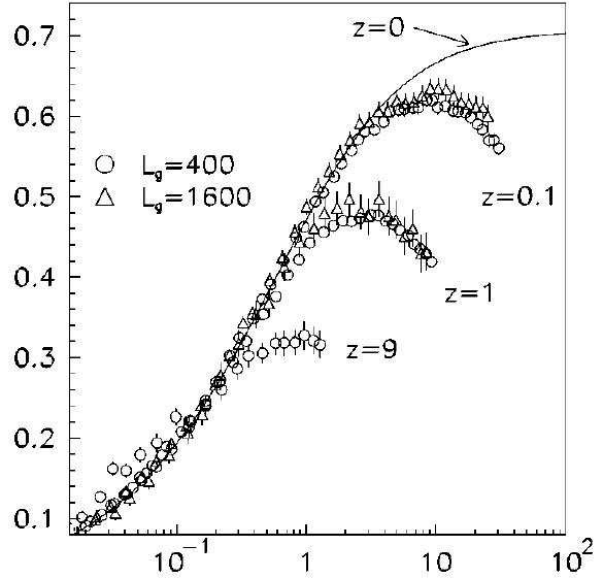}
   \caption{The ZFC $\chi _{ag} (t,s)$ is plotted versus the
            the ratio $[L(t)-L(s)]/L(s)$, for $L_g=400,1600$,
            and $z=0.1,9$. The solid line is the exact result for $z\to 0$.}
   \label{figchi}
\end{center}
\end{figure}

\begin{figure}[htpb]
\begin{center}
   \includegraphics[width=8cm]{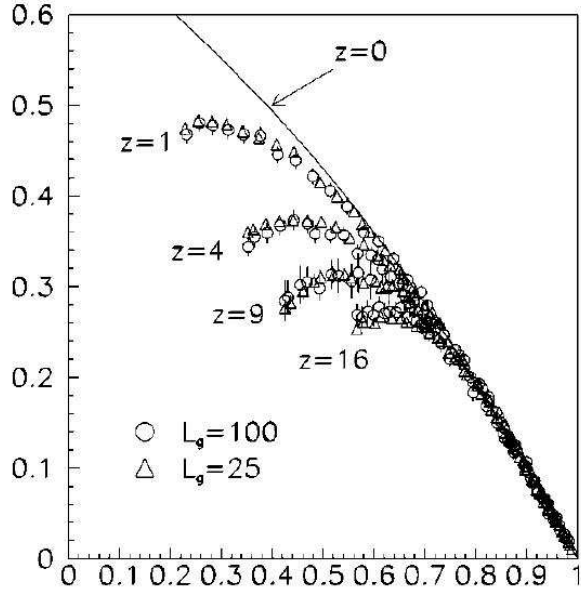}
   \caption{The ZFC $\chi _{ag} (t,s)$ is plotted versus the
            autocorrelation function $C(t,s)$, for $L_g=25,100$,
            and $z=1,4,9,16$. The solid line is the exact result for $z\to 0$.}
   \label{figchidic}
\end{center}
\end{figure}

Let us consider the case of the $d=1$ random field Ising model
(RFIM).
In the presence of a quenched random field, 
domain walls perform random walks in a
random potential of the Sinai type and the average
domain size $L(t)$ behaves as the root mean square displacement of the
random walker\cite{Fisher01}. The typical potential
barrier encountered by a walker after traveling a distance $l$ is order of 
$\sqrt{l\sigma_h}$ where $\sigma_h$ is the
variance of the random field. Hence, there exists a
characteristic length $L_g= T^2/\sigma_h$ representing the 
distance over which 
potential barriers are of the order of magnitude of thermal energy. 
For displacements much less than $L_g$ diffusion takes place in a flat
landscape like in the pure system and 
$L(t) \sim t^{1/2}$. For displacements much greater than $L_g$, instead, 
one finds the Sinai\cite{Sinai82} diffusion law $L(t) \sim (\ln t)^2$. 
The response function obeys\cite{Corberi4} the scaling relation (Fig.~\ref{figchi})
\be
T \chi_{\text{ag}}(t,s,L_g)= \widetilde{\chi}
\left ( \frac{L(t)}{L(s)},z \right )
\label{3.2.1}
\ee 
where $z=L(s)/L_g$.
For $z=0$ the form of the response function 
for  the pure system is recovered.
With $z>0$ there is a crossover. The pure case behavior holds for
$L(t)-L(s)\ll L_g $, while for $L(t)-L(s) > L_g $ the response 
levels off and then decreases. This is clearly displayed also in the 
plot (Fig.~\ref{figchidic}) against the autocorrelation function.  
Looking at the effective response of a
single interface $\chi^s(t,s)$  
one finds
$\chi^s(t,s,L_g) = L(s)
\widetilde{\chi}^s
\left ( \frac{L(t)}{L(s)},z \right )$
with the scaling function displaying the behavior
\be
\widetilde{\chi}^s(x,z) \sim \left\{\begin{array}{ll}
	x\widetilde{\chi}(x,z=0)       & \mbox{, for $x-1 \ll 1/z$} \\  
		\sqrt{x}        & \mbox{, for $x-1 \gg 1/z$.}
                   \end{array}                                        
               \right.
\label{3.9.2}
\ee
From this follows ${\chi}^s(t,s) \geq \rho_I^{-1}(t)$ 
in the preasymptotic regime and
${\chi}^s(t,s) \sim \rho_I^{-1/2}(t)$
in the asymptotic regime, which account for the crossover of the
response function in Figs.~\ref{figchi},\ref{figchidic} in terms of the balance between the rate
of growth of the single interface response and the rate of loss of
interfaces. Hence, for $z>0$ eventually $\widetilde{\chi}(x,z)$
vanishes and in the limit $z \rightarrow \infty$ one
expects
$\chi_{\text{ag}}(x,z=\infty) \equiv 0$.
Therefore, for any finite quenched random field the validity of Eq.~(\ref{4.3})
is restored.

\vspace{1cm}

\vspace{1cm}

\subsubsection{Failure by stochastic instability} \label{failstoc}

\begin{figure}[htpb]
\begin{center}
   \includegraphics[width=8cm]{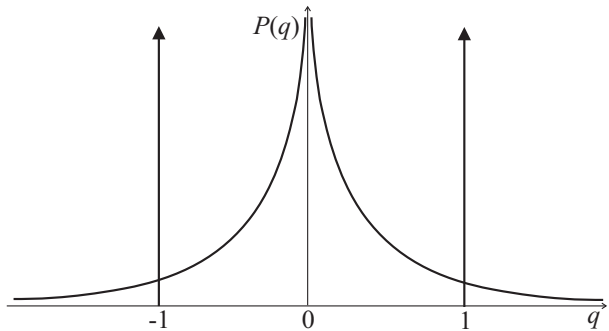}
   \caption{Overlap distribution for mean spherical model with
$M^2=1$. The arrows represent the $\delta$ functions of the
overlap distribution for the spherical model.}\label{overlap}
\end{center}
\end{figure}

An interesting example~\cite{Fusco}, where statics cannot be reconstructed from dynamics because the
third requirement of stochastic
stability is not satisfied, comes from the spherical model. More precisely,
one must consider in parallel the original version of the spherical model (SM) of Berlin and
Kac~\cite{Berlin} and the mean spherical model (MSM) introduced by Lewis and Wannier~\cite{Lewis}, with the
spherical constraint treated in the mean. These two models are equivalent above
but not below  $T_C$~\cite{Kac}. The low temperature states are quite different, with a bimodal
order parameter probability distribution in the SM case and a Gaussian distribution
centered in the origin in the MSM case.  The corresponding  overlap functions are also very
different~\cite{Fusco}. Considering, for simplicity, $T=0$ one has
\be
    P(q) = \left \{ \begin{array}{ll}
        \frac{1}{2} \left [ \delta (q-M^2) + \delta (q+M^2) \right ]
\qquad $for SM$ \\
        \frac{1}{\pi M^2}K_0(|q|/ M^2) \qquad $for MSM$
        \end{array}
        \right .
        \label{5.1}
        \ee
where $K_0$ is a Bessel function of imaginary argument
(Fig.\ref{overlap}).

However, after switching of an external field, one finds for {\it
both} models $\widetilde{P}(q) = \delta (q-M^2)$. This means that
stochastic stability holds for SM but not for MSM.

On the other hand, the relaxation properties are the same in the two models, both above and below
$T_C$ if the thermodynamic limit is taken before the $t \rightarrow \infty$ limit~\cite{Fusco}.
Then, the linear
response function is the same for both models and obeys Eq.~(\ref{2.9}) with
$a_{\chi}$ given by  Eq.~(\ref{3.1}), where $\delta$, $d_L$ and  $d_U$ are the same as for
the large $N$ model (Table I). Hence, we have that although Eq.~(\ref{4.3}) is satisfied for
both models, nonetheless statics and dynamics are connected only in the SM case,
where $\widetilde{P}(q) = \widehat{P}(q)$. Instead, this is not possible in the MSM case
where $\widetilde{P}(q) \neq \widehat{P}(q)$.

\section{Conclusions} \label{concl}

In this Article we have reviewed some recent progresses in the field of
non-equilibrium linear response theory. A first accomplishment is the
derivation of a generalization of the FDT for Markov processes
which allows the computation of the response function in terms of
correlation functions of the unperturbed system. This represents a great 
simplification particularly in numerical calculations, which are
usually computationally very demanding: The generalization of the FDT
allows a sensible speed and
precise numerical determination of the response function can be achieved. 
This quantity has been deeply investigated particularly in the field 
of slowly relaxing systems, because its relation with the autocorrelation
function represents a bridge between statics and dynamics.

Phase-ordering systems can be regarded as the simplest instance of 
aging systems, where the behavior of the response function
can be more easily investigated. 
In this context, a partial understanding has been achieved by matching
the results of numerical simulations with the outcomes of solvable models
and approximate theories, showing that the
scaling properties of the response function are non-trivial.
In particular, Eq.~(\ref{2.800}) is obeyed with $a_\chi $ depending
on dimensionality through the phenomenological formula~(\ref{3.1}), 
which is found to be consistent with all the cases 
considered in the literature and, for the scalar case is supported by 
an argument based on the roughness properties of the interfaces.
The dependence of $a_\chi $ on dimensionality is such that 
it vanishes at the lower critical dimension. This implies
an asymptotic finite contribution of the aging part of the
response function which invalidates the connection between statics
and dynamics. Phase ordering therefore offers examples
where a replica symmetric low temperature state
is compatible with a non trivial FDR which, therefore, 
cannot be used to infer the properties of the equilibrium state. 

This whole phenomenology is not adequately captured by the existing approaches
to phase-ordering. Theories based on the GAF method, originally 
introduced by Otha, Jasnow and Kawasaki~\cite{ojk}, provide the 
phenomenological formula~(\ref{3.1}) but with a wrong value 
$d_U=2$~\cite{Berthier99,Corberi2001}. This discrepancy is not removed
using a perturbative expansion~\cite{Mazenko2004} developed to improve over
the GAF approximation. Next to these theories, it is of much interest
the approach by Henkel {\it et al.}~\cite{Henkel2001}, based on the conjecture
that the response function transforms covariantly under the group
of local scale transformations. This {\it ansatz}, however, fixes the form
of the scaling function in Eq.~(\ref{2.800}) but not the exponent $a_\chi$
which remains insofar an undetermined quantity. A first principle theory
for the complete description of the behavior of the linear response function
in phase-ordering systems may represent a pre-requisite for understanding
the behavior of more complex systems, like glasses and spin glasses.
However, despite some progresses of a specific character, such a theory
is presently still lacking.

\end{document}